ORIGINAL ARTICLE

# A priori knowledge-free fast positioning approach for BeiDou receivers

Sihao Zhao[1] · Xiaowei Cui[1] · Mingquan Lu[1]

**Abstract** A Global Navigation Satellite System (GNSS) receiver usually needs a sufficient number of full pseudorange measurements to obtain a position solution. However, it is time-consuming to acquire full pseudorange information from only the satellite broadcast signals due to the navigation data features of GNSS. In order to realize fast positioning during a cold or warm start in a GNSS receiver, the existing approaches require an initial estimation of position and time or require a number of computational steps to recover the full pseudorange information from fractional pseudoranges and then compute the position solution. The BeiDou Navigation Satellite System (BDS) has a unique constellation distribution and a fast navigation data rate for geostationary earth orbit (GEO) satellites. Taking advantage of these features, we propose a fast positioning technique for BDS receivers. It simultaneously processes the full and fractional pseudorange measurements from the BDS GEOs and non-GEOs, respectively, which is faster than processing all full measurements. This method resolves the position solution and recovers the full pseudoranges for non-GEOs simultaneously within 1 s theoretically and does not need an estimate of the initial position. Simulation and real data experiments confirm that the proposed technique completes fast positioning without a priori position and time estimation, and the positioning accuracy is identical with the conventional single-point positioning approach using full pseudorange measurements from all available satellites.

✉ Sihao Zhao
zsh_thu@tsinghua.edu.cn

1 Department of Electronic Engineering, Tsinghua University, Beijing 100084, China

## Introduction

Similar to other global navigation satellite system (GNSS) receivers, a BeiDou Navigation Satellite System (BDS) receiver acquires satellite signals, synchronizes pseudorandom codes and navigation message frames, and decodes navigation data after it starts. According to the principles of GNSS, the receiver cannot complete its position fix without valid ephemerides and full pseudorange measurements from at least 4 satellites (Kaplan and Hegarty 2006). After the pseudorandom noise (PRN) code is tracked and synchronized, only the sub-code length (e.g., 1 ms for BDS B1I ranging code) fractional pseudorange information can be obtained which lacks the unknown number of integer code lengths from its full counterpart. The recovery of the full pseudorange measurements is the bottleneck for the time to first fix after ephemerides become available to the receiver. When having finished frame synchronization, the full pseudorange measurements can be only determined after having obtained the second of week (SOW) information periodically broadcast in the navigation message. This usually takes at least 6 s for non-geostationary earth orbit (non-GEO) BDS satellites which is identical to Global Positioning System (GPS) (Navstar GPS Joint Program Office 2006; CSNO 2013b). To shorten this time, the technique proposed by van Diggelen (2002, 2009) can be applied; however, it requires that the initial bias caused by position and time estimation is less than 150 km. Sirola and Syrjarinne (2002, Sirola 2006) propose an exhaustive searching method which requires initial estimation boundaries and heavy computations. The method in Jing et al. (2015) uses measurements from geostationary earth orbit (GEO) satellites to first obtain an approximate position estimate, and then recovers full pseudoranges for non-GEOs and finally computes the user position, which is an application of van Diggelen (2002, 2009) in BDS with a pre-computation step to satisfy the initial constraint.

We first analyze the characteristics of BDS constellation and its GEO navigation message. Then, a fast positioning technique taking advantage of the high data rate feature of the GEO broadcast message is proposed. It simultaneously processes full pseudorange measurements from at least 4 GEOs and fractional pseudorange measurements from at least 1 non-GEO and computes the position and time solutions as well as recovers full pseudorange measurements for non-GEOs without any constraint on priori position estimation. Simulated constellation and pseudorange data as well as real BDS measurement data are used to verify the proposed algorithm. The results show that the method successfully yields position solutions with full and fractional pseudorange information from GEOs and non-GEOs, respectively, which can reduce the time to first fix to less than 1 s theoretically. The proposed technique also provides an identical positioning accuracy with conventional methods using full pseudorange measurements from all available satellites and has no requirements for initial estimation of position and time. The proposed technique can be easily implemented in receivers and applied in a large area inside the BDS service region.

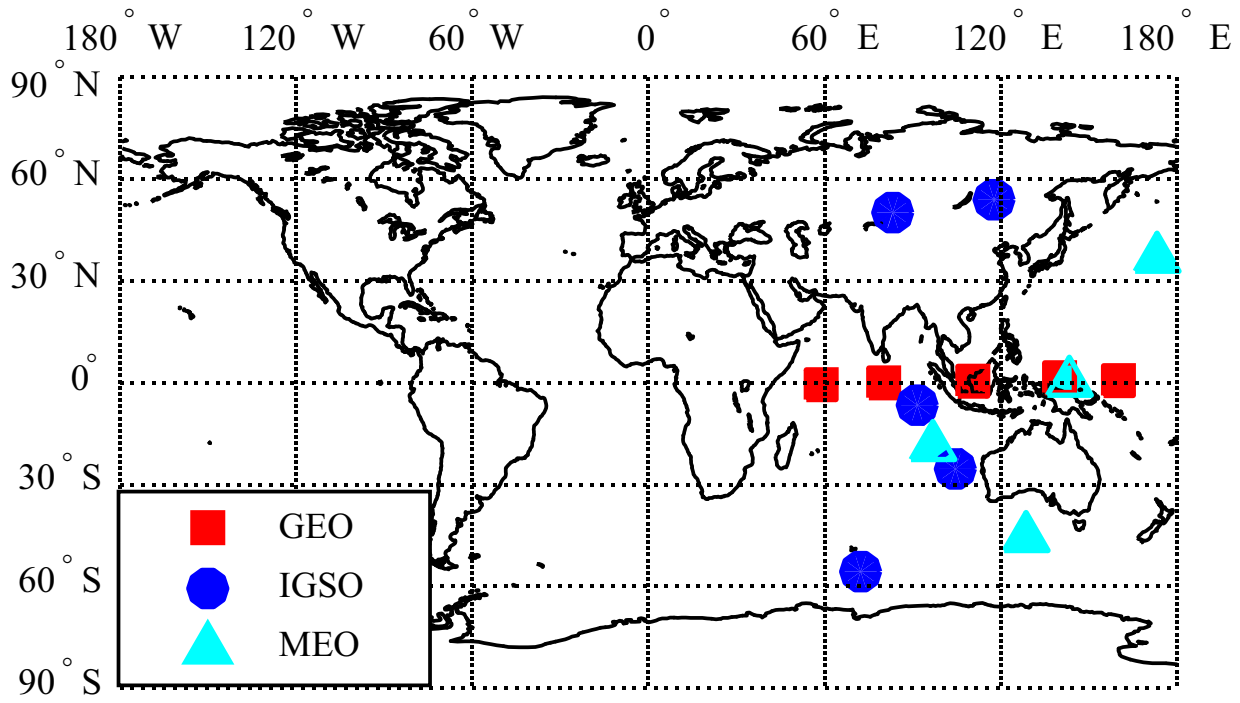

**Fig. 1** Snapshot of sub-satellite points of BDS constellation on 2015-05-19, 04:00:00 UTC

## Unique features of BDS

The current regional BDS has been offering service to Asia–Pacific area since the end of 2012. The space segment comprises 5 GEOs and 9 non-GEOs (5 inclined geostationary earth orbit satellites—IGSO and 4 medium earth orbit satellites—MEO) (CSNO 2013b). A snapshot of the distribution of BDS constellation is shown in Fig. 1, and their orbital elements at this moment is listed in Table 1. Although the signal of MEO 5 has not been received by the authors' receivers for some time as this work is under progress, for the completeness of the constellation, all the space vehicles announced in (CSNO 2013a) are included in the figure and table. The GEOs and IGSOs offer high elevation angle observation to the mid- to-high-latitude users in the regional service area which improves availability especially for canyon environments. The MEOs provide coverage beyond the typical service region as far as possible.

There are two types of BDS navigation messages called D1 and D2 for non-GEOs and GEOs, respectively. The data rate of the D1 navigation message for non-GEOs is 50 bps, and there is one SOW in each sub-frame consisting of 300 bits. As a result, it takes at least 6 s for the receiver to decode the first SOW and obtain the unambiguous full pseudorange measurement, and not to mention that extra time is always spent to check the reliability. However, for GEOs, the D2 data rate is 500 bps, and there is also one SOW in each 300-bit sub-frame. Therefore, only 0.6 s is needed to decode the first SOW theoretically and yield the full pseudorange measurement, which is ten times as fast as for non-GEOs.

**Table 1** BDS constellation orbital elements (UTC 2015-05-19, 04:00:00)

| Satellite ID | Semi-major axis (km) | Eccentricity | Inclination (°) | RAAN (°) | Argument of perigee (°) | True anomaly (°) |
|---|---|---|---|---|---|---|
| G1 | 42,167.046 | 0.000385 | 1.660 | 14.091 | 166.081 | 256.057 |
| G3 | 42,168.221 | 0.000371 | 1.580 | 31.719 | 312.433 | 62.552 |
| G4 | 42,165.839 | 0.000550 | 0.787 | 41.955 | 155.089 | 259.219 |
| G5 | 42,164.381 | 0.000110 | 1.158 | 29.645 | 72.196 | 253.240 |
| G6 | 42,165.779 | 0.000097 | 0.216 | 47.928 | 22.396 | 306.033 |
| I1 | 42,157.559 | 0.004004 | 54.293 | 200.802 | 210.089 | 1.845 |
| I2 | 42,170.380 | 0.003126 | 53.938 | 319.904 | 196.870 | 255.448 |
| I3 | 42,144.000 | 0.002785 | 57.010 | 80.717 | 193.820 | 86.116 |
| I4 | 42,162.302 | 0.003603 | 54.608 | 203.045 | 203.863 | 344.311 |
| I5 | 42,166.206 | 0.002972 | 54.031 | 319.350 | 195.436 | 235.874 |
| M3 | 27,907.553 | 0.002132 | 55.838 | 79.515 | 207.483 | 152.594 |
| M4 | 27,905.796 | 0.002440 | 55.773 | 78.992 | 199.825 | 206.342 |
| M5 | 27,905.263 | 0.003448 | 54.728 | 199.113 | 170.282 | 33.123 |
| M6 | 27,905.161 | 0.001848 | 54.831 | 198.716 | 214.969 | 26.118 |

It is a natural thought that we can take advantage of the faster data rate in GEOs and use their full pseudorange measurements to calculate the position and time solution instantly to shorten the time to first fix. However, it should be noted that the geometry of using only GEOs is too poor, nearly collinear as shown in Fig. 1, to yield a solution with acceptable accuracy. One or more full pseudorange measurement from non-GEOs can definitely improve the satellite geometry but at least another 6 s should be spent to generate the full pseudorange for conventional receivers. Within 0.6 s, only an ambiguous fractional pseudorange measurement, e.g., sub-millisecond value if PRN code is synchronized or sub-twenty-millisecond value if navigation bit is synchronized, can be obtained for non-GEOs. If this information can be utilized along with the full measurements from GEOs, a position solution with similar or identical accuracy to conventional single-point positioning approaches can be possibly obtained. The time cost can be potentially reduced to within 1 s theoretically which is much faster than using all full measurements. Based on this motivation, our fast positioning algorithm, using 4 or more full pseudorange measurements from GEOs and 1 or more fractional measurements from non-GEOs and no prior position estimation, is developed and tested in the following sections.

## Fast positioning technique without a priori position constraint

In this section, the conditions needed to apply the proposed approach are discussed after which the mathematical models and the algorithm procedures are presented.

### Basic conditions

In order to obtain the full pseudorange and the position information of any satellite, the ephemerides must be available either through a recently (typically within one hour age for BDS) stored navigation data or an assistant approach from communication links (e.g., Internet or mobile service). Assuming this condition, the user receiver can benefit from the proposed fast positioning technique. Therefore, it is assumed that the receiver has stored the valid ephemerides for all BDS satellites which can be satisfied for a commonly used BDS receiver even if it experiences a restart. As analyzed in the previous section, the full pseudorange measurements for all GEOs can be obtained after frame synchronization which costs at least 0.6 s theoretically. Even if more time is needed to confirm a reliable synchronization, this time is still much shorter than the length of a 6 s data sub-frame for non-GEOs. Similar to the processing of GEO navigation data, theoretically, at least 6 s is needed to obtain a full pseudorange measurement for non-GEOs. During the period while frame synchronization for 4 GEOs is completed but is still in progress for non-GEOs, only the sub-millisecond (or sub-twenty-millisecond) fractional pseudorange measurements for non-GEOs which equals the full measurement modulo the product of light speed and 1 ms (or 20 ms), are available. The aim of our algorithm is to give single-point positioning results with these imperfect measurements during this period, which is much shorter than 6 s.

### Mathematical models

The full and fractional pseudorange measurements are modeled as

$$z_i = |\mathbf{X}_i - \mathbf{x}_r| + b + \varepsilon_i \tag{1}$$

$$z_j = \left|\mathbf{X}_j - \mathbf{x}_r\right| + b - N_j c_T + \varepsilon_j \tag{2}$$

where $z_i$ and $z_j$ are the full and fractional pseudorange measurements for GEOs and non-GEOs, respectively, $\mathbf{X}_i$ and $\mathbf{X}_j$ are the three-dimensional satellite position vectors for GEO and non-GEO, respectively, $i \in \{1,\ldots,n\}$ and $j \in \{n{+}1,\ldots,n{+}m\}$ are indices of GEOs and non-GEOs, respectively, $\mathbf{x}_r$ is the three-dimensional receiver position, $b$ is the receiver clock bias which is identical for all measurements, $N_j$ is the pseudorange integer ambiguity which is the integer part of the pseudorange divided by $c_T$, $c_T$ is a constant distance the light travels in the time period $T$ (1 ms for a code chip and 20 ms for a navigation bit), and $\varepsilon_i$ and $\varepsilon_j$ are measurement noises.

The unknowns to be resolved include the receiver position, clock bisas, all pseudorange ambiguities for non-GEOs and is expressed as,

$$\mathbf{X} = \left[\mathbf{x}^T, \mathbf{N}^T\right]^T = [x, y, z, b, N_1, N_2\ldots, N_m]^T \tag{3}$$

where $\mathbf{x} = [x, y, z, b]^T$ and $\mathbf{N} = [N_1, N_2, \ldots, N_m]^T$.

The positions of satellites are computed based on the ephemerides and time of transmission. The satellite position computation algorithm is commonly found in literature such as CSNO (2013b). The time of transmission for GEOs and non-GEOs is presented as

$$t_{t,i} = t_{r,i} - z_i/c - t_{e,i} \tag{4}$$

$$t_{t,j} = t_{r,j} - \left(z_j + N_j c_T\right)/c - t_{e,j} \tag{5}$$

where $t_{t,i}$ and $t_{t,j}$ are the time of transmission for the $i$th GEO and the $j$th non-GEO, respectively, $t_{e,i}$ and $t_{e,j}$ are satellite clock corrections calculated from the navigation message, and $c$ is the speed of light.

An iterative technique is used to solve the unknowns in (3) when at least 4 full measurements in (1) and at least 1 fractional measurement in (2) form an equation set. The incremental vector for each iterative step is defined as

$$\delta\mathbf{X} = [\delta x, \delta y, \delta z, \delta b, \delta N_1, \ldots \delta N_m]^T \tag{6}$$

The measurement residuals for every step are given in

$$\delta z_i = z_i - \left|\hat{\mathbf{X}}_i - \hat{\mathbf{x}}_r\right| - \hat{b} - \hat{d}_i \tag{7}$$

$$\delta z_j = z_j - \left|\hat{\mathbf{X}}_j - \hat{\mathbf{x}}_r\right| - \hat{b} + \hat{N}_j c_T - \hat{d}_j \tag{8}$$

where "^" represents the estimates at the current step and $\hat{d}$ denotes other estimated errors such as ionosphere and troposphere delays. The vector form of measurement residuals is written in

$$\delta Z = [\delta z_1, \ldots, \delta z_n, \delta z_{n+1}, \ldots, \delta z_{n+m}]^T \tag{9}$$

The least-squares solution of the incremental vector is

$$\delta\mathbf{X} = \left(\mathbf{H}^T\mathbf{H}\right)^{-1}\mathbf{H}^T\delta\mathbf{Z} \tag{10}$$

The design matrix $\mathbf{H}$ in (10) is defined as

$$\mathbf{H} \doteq \begin{bmatrix} -\mathbf{e}_1^T & 1 & & & \\ \vdots & \vdots & & \mathbf{O}_{n\times m} & \\ -\mathbf{e}_n^T & 1 & & & \\ -\mathbf{e}_{n+1}^T & 1 & -c_T & & \\ \vdots & \vdots & & \ddots & \\ -\mathbf{e}_{n+m}^T & 1 & & & -c_T \end{bmatrix} \tag{11}$$

where $\mathbf{e}$ is the normalized line-of-sight (LOS) vector from the estimated position to the corresponding satellite.

Another matrix $\mathbf{D}$ is defined below. The eigenvalues of this matrix are used to examine if the algorithm can obtain the correct position, i.e., if the sum of the reciprocals of the eigenvalues exceeds a predefined threshold, the algorithm is unlikely to compute the correct solution and will be aborted. The principle and the selection of the threshold of this examination will be discussed in the usability section. The matrix is

$$\mathbf{D} \doteq \begin{bmatrix} -\mathbf{e}_1 & \cdots & -\mathbf{e}_n \\ 1 & \cdots & 1 \end{bmatrix} \begin{bmatrix} -\mathbf{e}_1^T & 1 \\ \vdots & \vdots \\ -\mathbf{e}_n^T & 1 \end{bmatrix} \tag{12}$$

It should be noted that the pseudorange ambiguity $N$ is an integer. Therefore, the updated unknowns should keep this integer feature in each step, and the update step is written as

$$\hat{\mathbf{X}}_{k+1} = \hat{\mathbf{X}}_k + [\delta x, \delta y, \delta z, \delta b, \mathrm{round}(\delta N_1), \ldots, \mathrm{round}(\delta N_m)]^T \tag{13}$$

where the operator "*round*" represents rounding to the nearest integer.

After several iterations, the elements related to position and clock in the incremental vector converge close to zero. The ambiguity-related incremental elements ($\delta N_1$,…, $\delta N_m$) stabilize at values that are much smaller than 0.5, i.e., all the rounded incremental integer ambiguities are zero and all the ambiguities converge to specific integers that can correctly recover the full pseudorange measurements for non-GEOs. This convergence can be expressed as

$$\left|\hat{\mathbf{X}}_{k+1} - \hat{\mathbf{X}}_k\right| < \theta \tag{14}$$

where $\theta$ is a predefined positive quantity close to zero.

At this stage, even though the position solution converges, it is still far from accurate. The reason is that the errors of the position solution are projected onto the ambiguity incremental elements which should strictly be zero due to its integer feature. These errors shall be corrected to obtain the final solution. The integer ambiguities of all fractional pseudorange measurements are found at this point. Therefore, these converged pseudorange integer ambiguities can be treated as known measurements and only the position and clock bias are regarded as unknowns. Thus, a new design matrix is

$$\mathbf{A} = \begin{bmatrix} -\mathbf{e}_1^T & 1 \\ \vdots & \vdots \\ -\mathbf{e}_n^T & 1 \\ -\mathbf{e}_{n+1}^T & 1 \\ \vdots & \vdots \\ -\mathbf{e}_{n+m}^T & 1 \end{bmatrix} \tag{15}$$

and the corrected position and clock bias solution is given by.

$$\hat{\mathbf{x}}_k = \hat{\mathbf{x}}_k + \left(\mathbf{A}^T\mathbf{A}\right)^{-1}\mathbf{A}^T\delta\mathbf{Z} \tag{16}$$

At this point, both the final position solution and the recovered full pseudorange measurements are computed. Compared with the widely adopted iterative positioning technique, the proposed algorithm has similar iterative structure and does not add much computation burden to the existing approach, so it is easy to be implemented in a receiver. It also can be seen that if full pseudorange measurements are available for non-GEOs, the incremental ambiguities in every step remains zero which indicates that the algorithm can also work under the conventional all full pseudorange case.

## Algorithm procedure

According to the mathematical models presented in the previous section, the procedure of the fast positioning algorithm is depicted in Fig. 2 and summarized as follows.

1. Initialization
   All the unknowns in (3) are set to zero. At least 4 full and at least 1 sub-millisecond fractional pseudorange measurements from GEOs and non-GEOs, respectively, are obtained. The ephemerides of all the satellites are ready for satellite position determination in next steps.
2. Iteration
   *Time of transmission calculation* Time of transmission is calculated based on (4), (5) and clock correction algorithm in CSNO (2013b) for GEOs and non-GEOs. *Satellite position calculation* Satellite positions are determined with the time of transmission and satellite ephemerides using the algorithm described in CSNO (2013b).

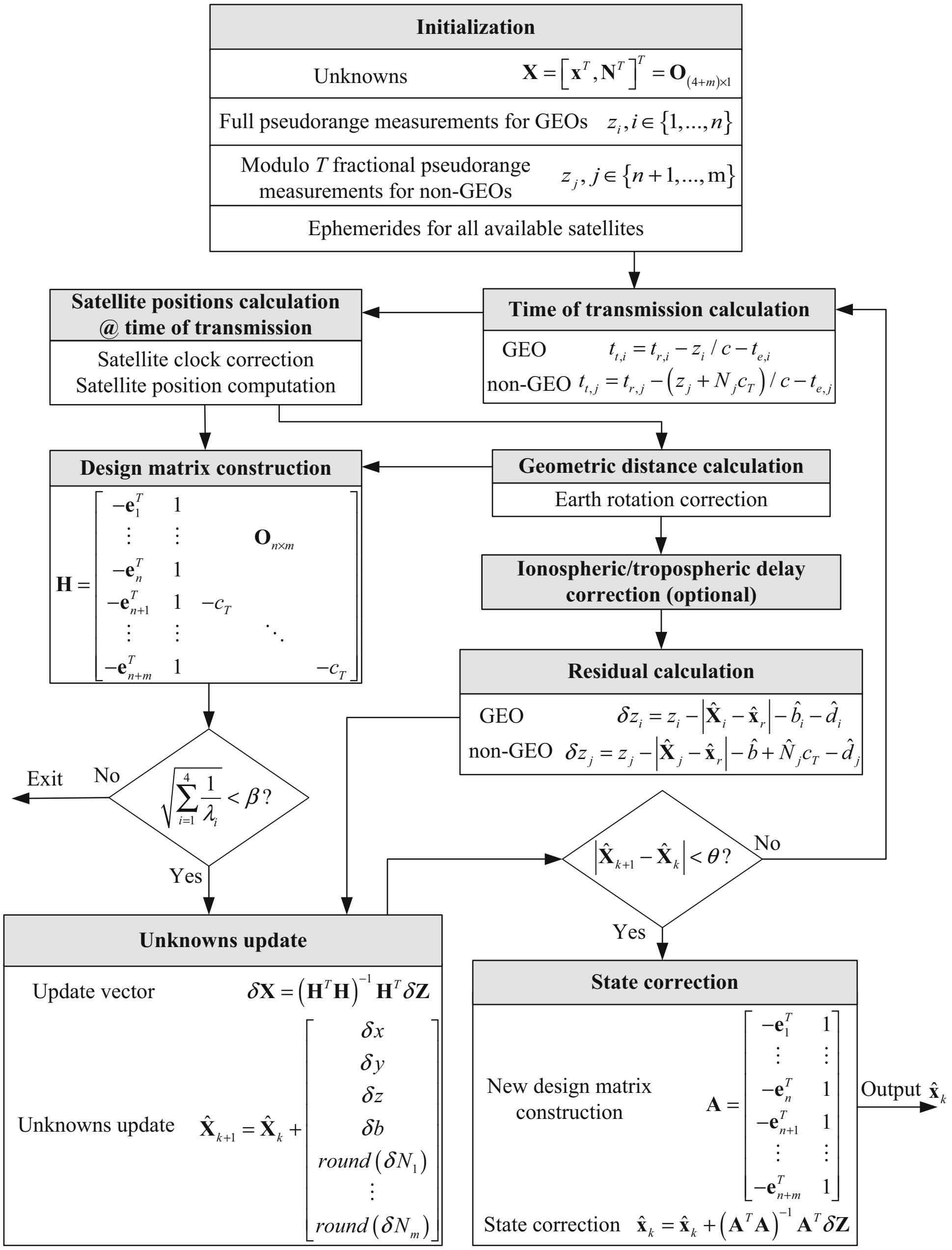

**Fig. 2** BDS fast positioning algorithm block diagram with data flow and mathematical models

*Geometric distance calculation* Distance between the estimated user position and every satellite is computed, and the correction for earth rotation during the signal travel time is applied according to Borre et al. (2007).
*Residual calculation* Measurement residuals are calculated based on (7) and (8) for GEOs and non-GEOs. The correction of ionospheric and tropospheric errors (Saastamoinen 1972; Klobuchar 1987) can be applied if necessary. These errors do not affect the performance of the proposed technique and are mentioned here for completeness of the positioning algorithm.
*Design matrix construction* The designed matrix is important to compute the corrections for every iterative step, and it is formed by (11).
*Eigenvalue test* The sum of the reciprocals of the eigenvalues of matrix **D** as defined in (12) is tested. If it exceeds a predefined threshold, then exit the algorithm and report a failure. Otherwise, the algorithm continues.
*Unknowns update* The correction vector for the current iteration is determined by (10), and the unknowns are updated by (13).
*Conditional Judgment* This module checks the convergence of the iteration. If the norm of the incremental vector is sufficiently small or the number of iteration exceeds a preset maximum value, the algorithm proceeds to 3) to yield the final solution. Otherwise, it goes back to 2) to resume iteration.

3. State correction
The new design matrix is first constructed by (15), and then the corrected position and clock solution are output by (16).

## Usability of the proposed algorithm

From the algorithm procedure presented previously, the correct recovery of pseudorange integer ambiguities as given in (13) is one of the key parts to the algorithm. It should be noted that this resolution of the pseudorange integer ambiguity is very similar to the case in carrier cycle integer ambiguity resolution that is extensively studied in the high-precision positioning area. The difference is that for the BDS D1 broadcast message, the code chip length and navigation bit length are about 300 km and 6000 km, respectively, while the carrier cycle is less than 20 cm for B1I signal (CSNO 2013b). Hence, we can use a simple rounding operation and obtain the correct integer ambiguity if the uncertainty of the estimated pseudorange, or "float" ambiguity as commonly referred to in carrier phase based techniques, is smaller than half of the chip length, i.e., 150 km; the code chip case is considered and tested rather than the navigation bit case because it is more stringent. In the proposed algorithm, we rely on the observation of the GEOs to resolve the non-GEOs' ambiguities. Therefore, in order to obtain the correct full pseudorange for one non-GEO, its user-to-satellite LOS error caused by the GEOs pseudorange errors projected onto the LOS vector shall be smaller than 150 km as given by

$$\left|\sum_{i=1}^{n}\left(\delta\rho_i \mathbf{e}_i^T \mathbf{e}_j^T\right)\right| < \alpha \tag{17}$$

where $\delta\rho_i$ is the random range error for the $i$th GEO, $\mathbf{e}$ is the normalized LOS vector from the receiver to the satellite, $i$ and $j$ are the indices of GEOs non-GEOs, respectively, and $\alpha$ is the half cycle threshold (150 km for the code case).

Furthermore, it will be simpler and easier if we can connect the geometric dilution of precision (GDOP) of the visible GEOs to the estimated non-GEO pseudorange error as expressed by

$$\begin{aligned}\left|\sum_{i=1}^{n}\left(\delta\rho_i \mathbf{e}_i^T \mathbf{e}_j^T\right)\right| &\le \left|\sum_{i=1}^{n}\delta\rho_i \mathbf{e}_i^T\right|\left|\mathbf{e}_j^T\right| = \left|\sum_{i=1}^{n}\delta\rho_i \mathbf{e}_i^T\right| \\ &= |\delta\mathbf{P}| \le \text{GDOP} \times \max_{i\in\{1,2,\cdots,n\}}(|\delta\rho_i|)\end{aligned} \tag{18}$$

where $\delta\mathbf{P}$ is the user position error vector. Unless otherwise specified, all GDOPs in this paper represent the GDOP of the visible GEOs. If the last product of (18), i.e., GDOP $\times$ $\max_{i\in\{1,2,\cdots,n\}}(|\delta\rho_i|)$ is smaller than the threshold $\alpha$, the criterion given by (17) can be replaced by a stricter one written as

$$\text{GDOP} \times \max_{i\in\{1,2,\cdots,n\}}(|\delta\rho_i|) < \alpha \tag{19}$$

It is further known from textbooks that GDOP of the visible GEOs is defined as the square root of the trace of matrix **D** (defined in 12), and it is also connected to its eigenvalues as

$$\text{GDOP} = \sqrt{\text{trace}\left(\mathbf{D}^{-1}\right)} = \sqrt{\sum_{i=1}^{4}\frac{1}{\lambda_i}} \tag{20}$$

where $\lambda$ is the eigenvalue of **D**. Based on (20), the test in (19) is equivalent to the test for GDOP or the eigenvalues of **D** as given in

$$\text{GDOP} = \sqrt{\sum_{i=1}^{4}\frac{1}{\lambda_i}} < \beta = \frac{\alpha}{\max_{i\in\{1,2,\cdots,n\}}(|\delta\rho_i|)} \tag{21}$$

where $\beta$ is the predefined threshold.

Using (21), we can determine whether or not the algorithm can obtain the correct position solution before computing the inverse matrix, i.e., if the square root term is smaller than $\beta$, the algorithm can compute the pseudorange ambiguities and the position solution successfully. The selection of $\beta$ must be taken care of. According to Kaplan and Hegarty (2006) and CSNO (2013a), the estimated $3\sigma$ of the user equivalent range error of BDS is less than 50 m, which is a conservative and strict estimate. Consequently, $\beta$ can be set to 3000 (150 km/50 m) for the code synchronized case.

In the following section, tests are conducted to find in what areas the algorithm can be applied and to verify the correctness of the GDOP (or the eigenvalue) test criterion.

## Experiment

Simulated and real data tests are used in this section. The setting of the simulation is first presented. Then, simulated constellation during an 8-day period is built based on real orbit elements and is used to test the usability of the proposed algorithm. The pseudorange measurements are then generated in the same simulation case to test the positioning performance. Finally, real BDS measurement data are preprocessed to generate the fractional pseudorange measurements and the data are input to the algorithm to test its performance under the real environment.

### Simulation settings

From the algorithm procedure and analysis presented above, at least 4 full pseudorange measurements must be obtained to initialize the algorithm. These are not limited to GEO measurements, but can be replaced by non-GEO measurements which indicates the generality and potential of extension of the algorithm. However, in typical cases, e.g., a restart of a BDS receiver, the full measurements from GEOs can be obtained earlier than from the non-GEOs due to the faster data rate of GEO broadcast messages as explained above. Therefore, only the cases with full measurements from GEOs and fractional ones from non-GEOs are considered and tested in this section. According to CSNO (2013a), the vertical coverage region of BDS satellites ranges from the earth surface to an altitude of 1000 km above the earth surface. Tests inside this region where at least 4 GEOs are visible will demonstrate the actual performance of the proposed method.

The ground area and space area over the globe are sampled with 1 degree in latitude and longitude between points. The receiver is assumed to be placed at these points to perform positioning with BDS signals. At every point, given a predefined clock bias, the pseudorange measurements generated by the receiver are simulated. The measurements for the non-GEOs are then truncated to keep its sub-millisecond values to test the proposed method. The original simulated full measurements can be used to compute the user position with a conventional single-point positioning approach. The two results are then compared to evaluate the performance of the proposed technique. The parameter settings for the simulation are given in Table 2.

**Table 2** Simulation parameters

| Item | Value |
|---|---|
| Clock bias (s) | 5 |
| Pseudorange noise ($1\sigma$) (m) | 1.3 |
| Elevation cut-off angle (°) | 0 |
| Initial $X$, $Y$ and $Z$ position estimate (m) | 0 |
| Initial clock bias estimate (s) | 0 |
| GDOP (eigenvalue test) threshold | 3000 |

After having completed the frame synchronization or obtained at least one full pseudorange measurement from a GEO, the bias of the receiver clock can be restricted within 20 ms which is about the difference between the maximum and the minimum time of light travelling from a GEO to a receiver on the earth surface or at the altitude of 1000 km. Considering this, the clock bias is set to 5 s which is much more severe than the worst case. The pseudorange noise is set to 1.3 m ($1\sigma$) according to CSNO (2013a). The elevation cut-off angle is set to zero in order to include the very edge areas where 4 GEOs are visible. The estimates of the receiver position and the clock bias are initialized to zero so that there is no priori constraint for the proposed algorithm. The GDOP threshold, or the eigenvalue test threshold, is set to 3000 to determine whether to exit or to continue the algorithm as analyzed in the previous section.

### Simulation test: usability

First, the areas where 4 or more GEOs are visible is illustrated based on simulation. The orbital elements of the BDS satellites in Table 1 are used to create the simulation case. The outer red line in Fig. 3 depicts the ground area and the inner blue line shows the space area at the altitude of 1000 km above the earth that is covered by 4 or more BDS GEOs based on the constellation distribution given in Fig. 1. The surface area ranges between 59°E to 161°E and 78°S to 79°N, with an east–west span of about 11,400 km and a north–south span of 17,500 km, while the area at the altitude of 1000 km is slightly smaller, ranging from 60°E to 159°E and 76°S to 77°N with an east–west span of about 11,000 km and a north–south span of 17,100 km. These two areas are selected to test both the usability and the positioning performance of the proposed fast positioning algorithm.

Based on the usability analysis mentioned above, the GDOP shall be smaller than a threshold to allow the algorithm to yield the correct pseudorange ambiguities and position results. In this test, we aim to identify the area and time where the algorithm can obtain a correct position solution. The MEOs of the BDS constellation have a recursion period of 13 rotations in 7 days (CSNO 2013a). Accordingly, the length of the simulation is set to 8 days to include all possible satellite positions. Figure 4 shows the proportion of the area where GDOP is smaller than 3000 in the area where 4 or more GEOs are visible at every time epoch on both earth surface and the altitude of 1000 km. The maximum proportion is 1 which means the GDOP $<$ 3000 area is identical with the area covered by 4 or more GEOs. Sometimes, the proportion falls below 1, which is due to the movement of the GEOs that causes a significant increase in GDOP. The minimum proportion is 76.7 % for the earth surface and 75.9 % for the 1000 km altitude. Figure 5 illustrates this worst case in which the area of GDOP $<$ 3000 is smaller than the area covered by 4 or more GEOs in Fig. 3. Considering all the region and time simulated, the entire proportion is 99.4 % which means theoretically that the proposed algorithm can generate the correct position solution for most of the time and region when and where 4 or more GEOs are visible. The proportion results are listed in Table 3.

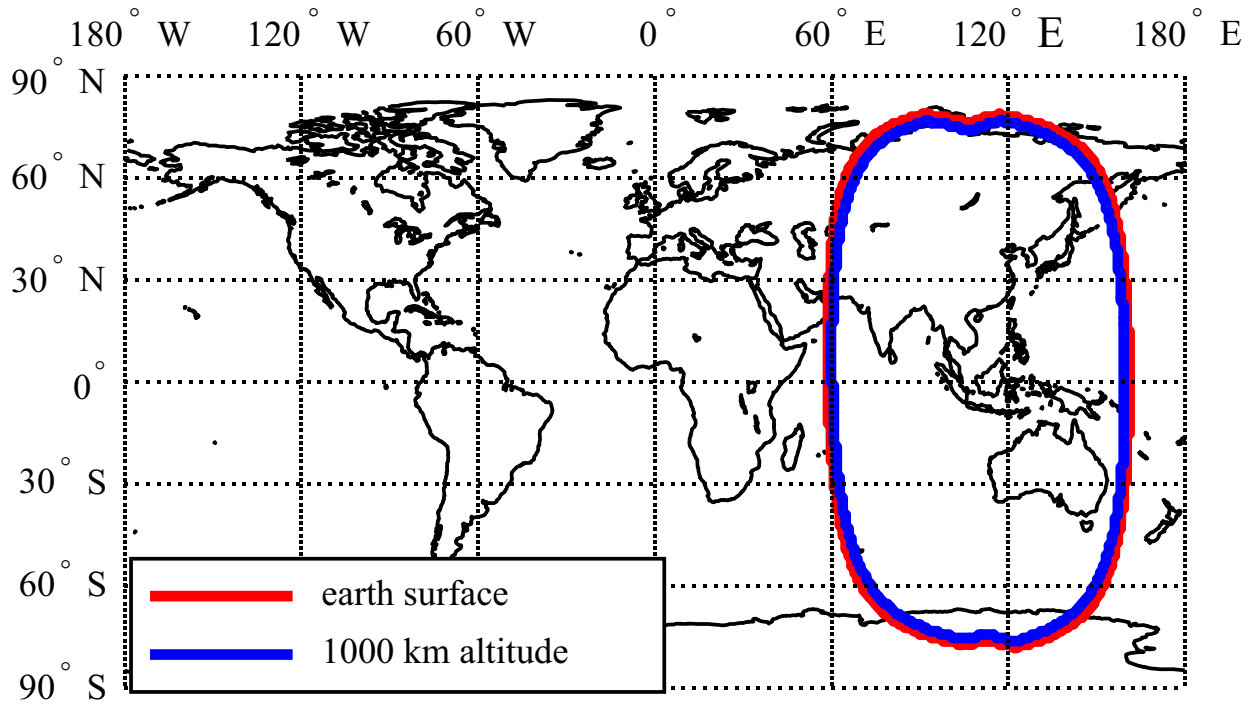

**Fig. 3** Coverage area by at least 4 GEOs on earth surface (*outer red*) and at the altitude of 1000 km (*inner blue*)

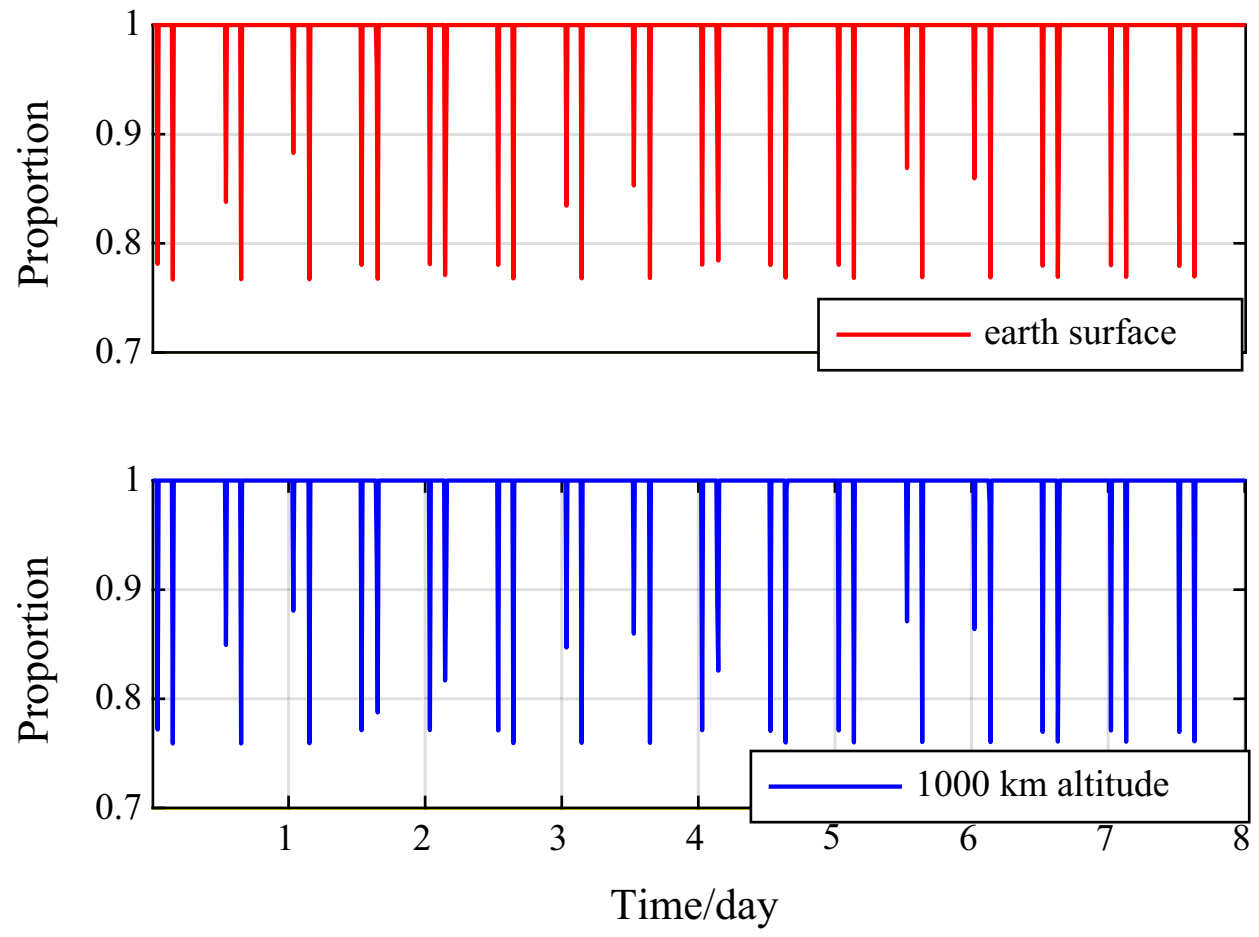

**Fig. 4** Proportion of points where GDOP (or eigenvalue test threshold) is smaller than 3000 among the points where the number of visible GEOs is no less than 4

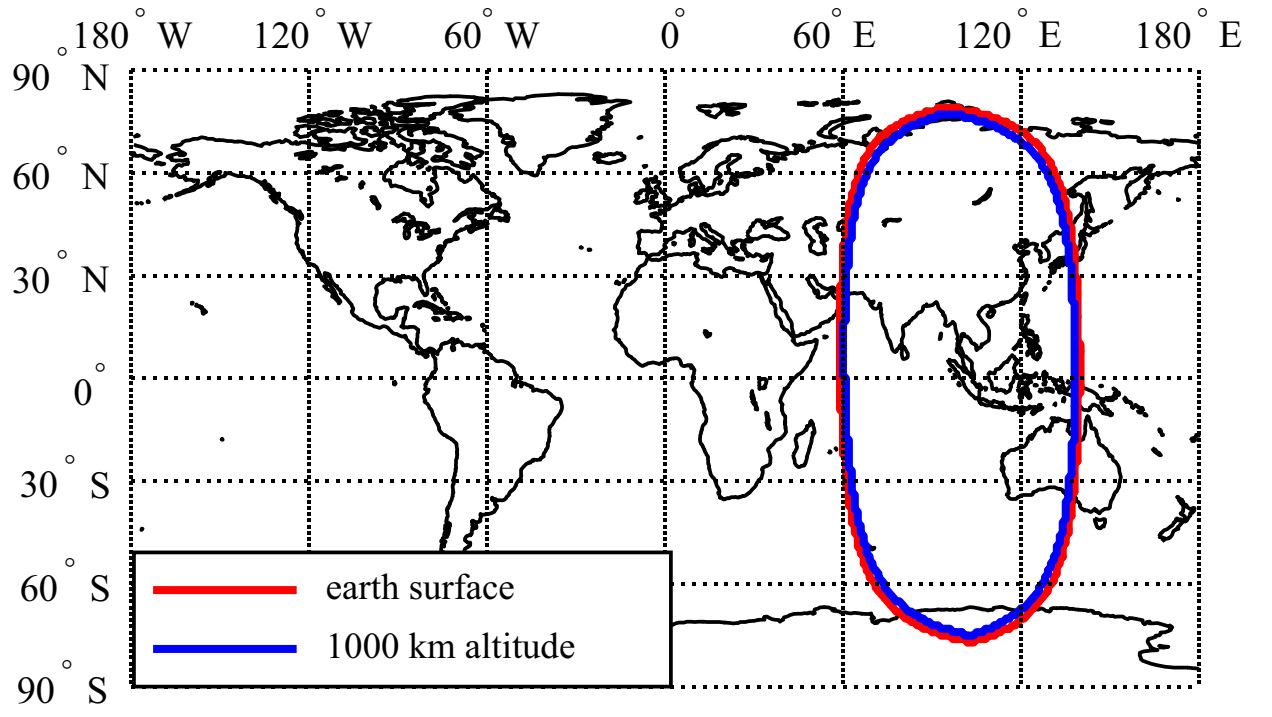

**Fig. 5** Area where the algorithm can offer correct position solutions in the worst case with the minimum proportion of the area where GDOP (or eigenvalue test threshold) is smaller than 3000 in the area with the number of visible GEOs $\geq$ 4

**Table 3** Proportion of the area with GDOP $<$ 3000 in the area with the number of visible GEOs $\geq$ 4

| | Max (%) | Min (%) | All simulation period (%) |
|---|---|---|---|
| Earth surface | 100 | 76.7 | 99.4 |
| 1000 km altitude | 100 | 75.9 | 99.4 |
| All area | 100 | 76.3 | 99.4 |

## Simulation test: positioning

The time epoch of one best case, having proportion equal to 1, from the usability simulation in the previous section is selected as the first simulation case in this part. In this test, at all points inside both the surface and the 1000 km altitude areas with 4 or more visible GEOs, the algorithm successfully resolves the pseudorange ambiguities for the non-GEOs and gives the correct positioning results. The root-mean-square errors (RMSE) of the three-axis position results in the earth-centered-earth-fixed (ECEF) frame compared with the simulated real position by both the proposed method and the conventional method with full measurements for the earth surface test are demonstrated in Fig. 6, and the corresponding results for the 1000 km altitude are similar and the figure is not included to save space. It can be observed that correct position results can be generated by the proposed algorithm for all points in the 4-GEO coverage area and the position error is identical with that of the conventional method with full measurements for both the earth surface and the 1000 km altitude cases. The worst case with the minimum proportion is also selected as a test case. The area where the algorithm

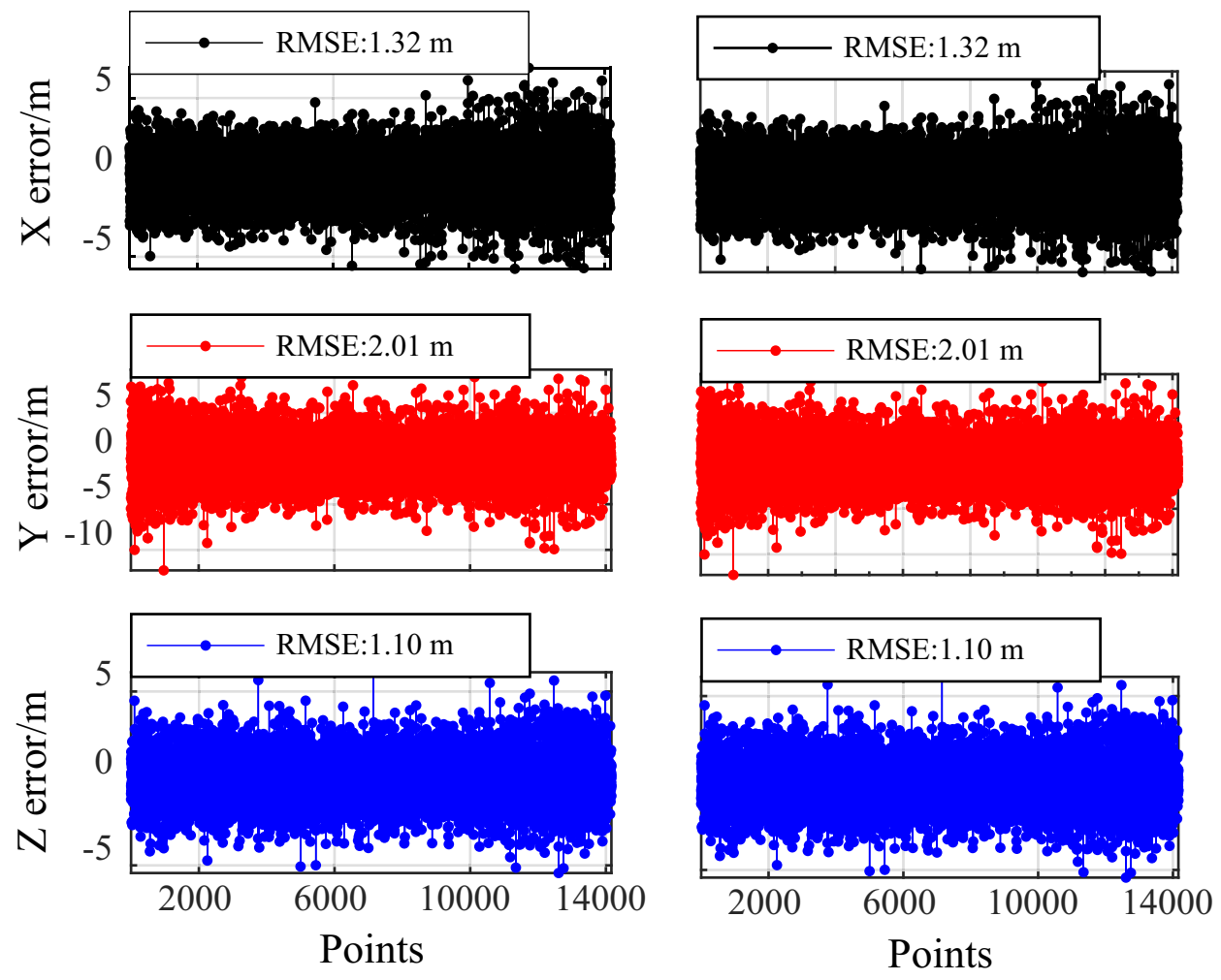

**Fig. 6** 3D position error for earth surface area (best case). *Left* the proposed algorithm using full and fractional measurements from GEOs and non-GEOs, respectively. *Right* the conventional method using full measurements from all satellites

generates correct positioning results is identical with the GDOP < 3000 area which proves the correctness of the eigenvalue test. The position RMSE which is not plotted here is also identical with that from the conventional method with full pseudorange measurements. Consequently, the usability and correctness of the proposed technique are validated. It can also be seen that the proposed fast positioning algorithm works inside the space where 4 or more GEOs are visible to the receiver.

## Real data test: positioning

Another experiment using field-collected real BDS measurements is conducted. A self-developed BDS receiver is used in this test and the antenna of which is placed statically on the rooftop of Weiqing Building, Tsinghua University, Beijing, China. Eight-hour BDS measurement data as well as the broadcast navigation message were collected starting at 08:30am (Beijing time), May 12, 2015. The visibility of all the satellites during this period is depicted in Fig. 7. It can be seen that the GEOs (PRN 1 to PRN 5) are visible throughout the dataset and some of the non-GEOs (PRN 6, 8, 9, 11, 12 and 14) experienced rising and setting during this period.

Similar to the previous simulated test, the pseudorange measurements of the non-GEOs are chopped to their sub-millisecond values. When processing the measurement data using either the proposed algorithm or the conventional approach, neither ionospheric nor tropospheric delay corrections are applied. The initial position estimate is set to zero which makes the proposed algorithm free of a priori constraint and is far from the true location. The position

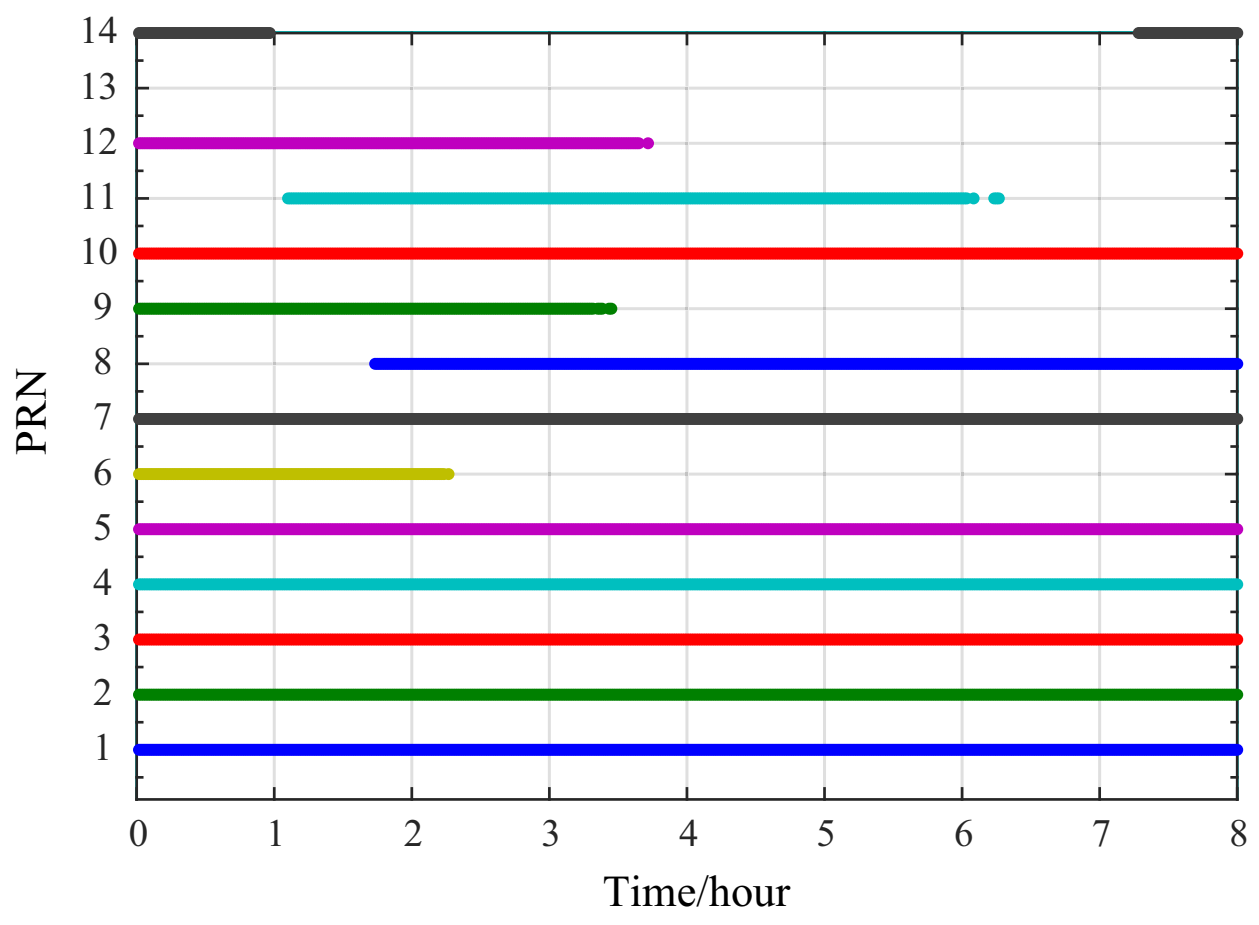

**Fig. 7** Satellite visibility during the real data test

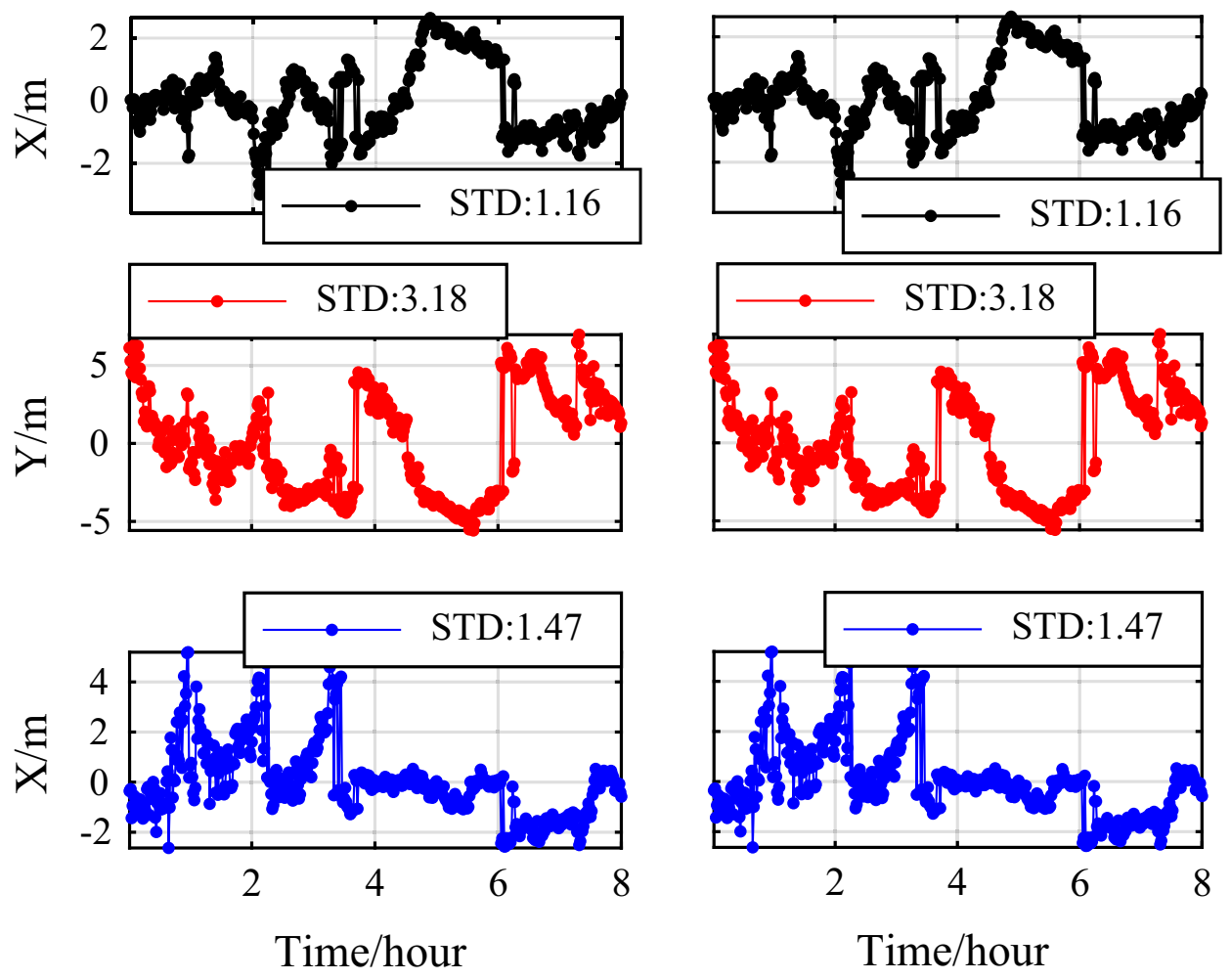

**Fig. 8** Position result with real data. The mean values of the results are subtracted for each axis. *Left* the proposed algorithm using full and fractional measurements from GEOs and non-GEOs, respectively. *Right* the conventional method using full measurements from all satellites

results and their standard deviations (STD) in ECEF frame from both methods are depicted in Fig. 8. For easy reading, the respective mean has been subtracted. It shows that the results of the proposed algorithm processing both full and fractional measurements are identical with that of the conventional method processing full measurements. This experiment verifies the effectiveness of the proposed method in the real environment.

It is worth mentioning that the proposed technique can also process full measurements from all satellites. In this case, the estimation of the pseudorange ambiguities simply converge to zero and the same positioning results can be obtained. Therefore the proposed technique also demonstrates its generality and substitutability over the conventional method.

## Conclusion and future work

Full pseudorange measurements from BDS non-GEOs need more time to be obtained than from GEOs. Based on this fact, a fast positioning algorithm that processes full measurements from GEOs and factional measurements from non-GEOs without a priori knowledge of position or time for BDS receivers is proposed. The principles, models, procedures and usability of the algorithm were presented. Simulated and real BDS measurement data were used to verify the usability and performance of the proposed method. Results show that in over 99 % of the time inside a vast space where 4 or more GEOs are visible in the BDS service region, the proposed algorithm yields position results with the same accuracy as that of the conventional method using full pseudorange measurements. Therefore, taking advantage of the fast data rate on BDS GEO broadcast signal, this method can be effectively applied in those cases when only 4 or more full pseudorange measurements of the GEOs and one or more fractional measurements of the non-GEOs are available to achieve fast positioning and reduce the time to first fix.

In the future, more tests with field-collected data from various locations in the BDS service area should be done to further validate the applicability of the proposed algorithm in real environments. Other techniques to reduce the dependence on the full GEO pseudorange measurements will be investigated. The algorithm will be implemented and improved in the self-developed BDS receivers for possible real-time applications.

**Acknowledgments** Many thanks to anonymous reviewers. Their insightful comments helped the authors to improve the manuscript. This work is supported by the China Postdoctoral Science Foundation (No. 2014M550732), China Civil Aviation Science and Technology Program (No. MHRD20140102), and the Key Program of the National Natural Science Foundation of China (No. U1333203).

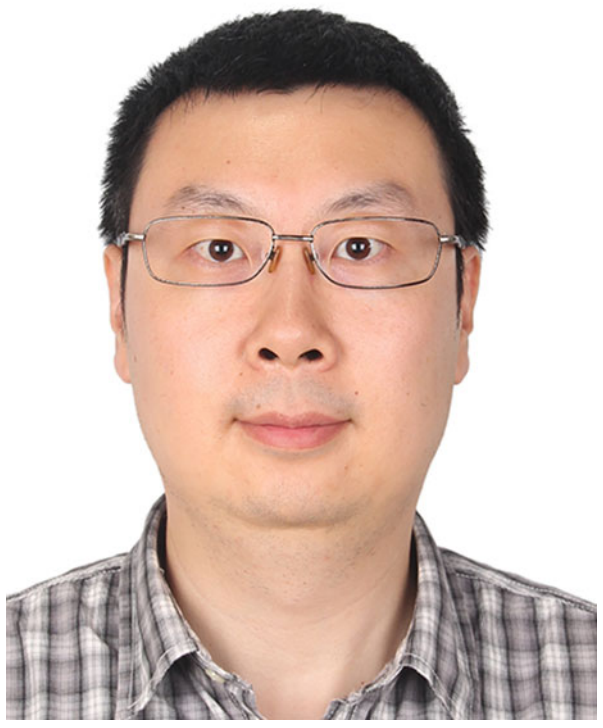

**Sihao Zhao** is an assistant research fellow at the Department of Electronic Engineering, Tsinghua University, Beijing, China. His research interests include GNSS signal processing algorithms, high-precision positioning techniques, indoor navigation systems as well as telecommunication systems design and verification. He received both his B.S. and Ph.D. degrees from the Department of Electronic Engineering, Tsinghua University.

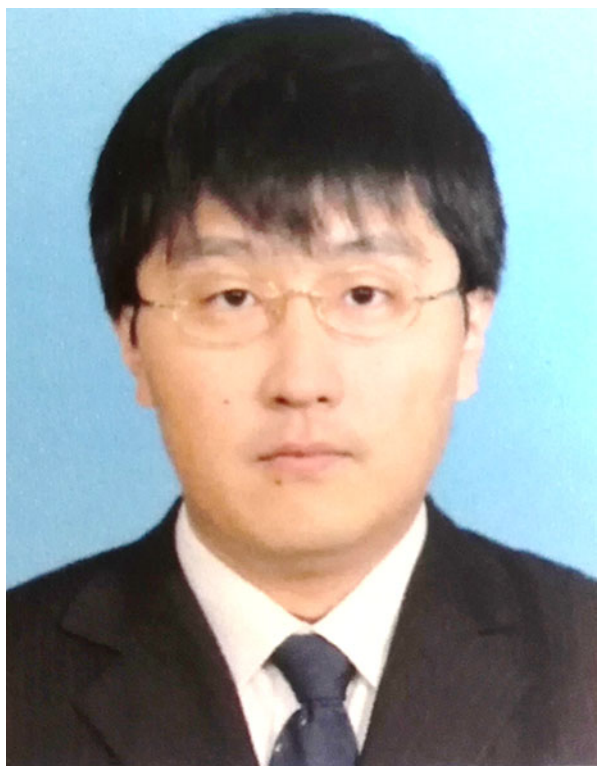

**Xiaowei Cui** is an associate professor at the Department of Electronic Engineering, Tsinghua University, Beijing, China. He is a member of the Expert Group of China BeiDou Navigation Satellite System. His research interests include robust GNSS signal processing, multipath mitigation techniques and high-precision positioning. He obtained both his B.S. and Ph.D. degrees in Electronic Engineering from Tsinghua University.

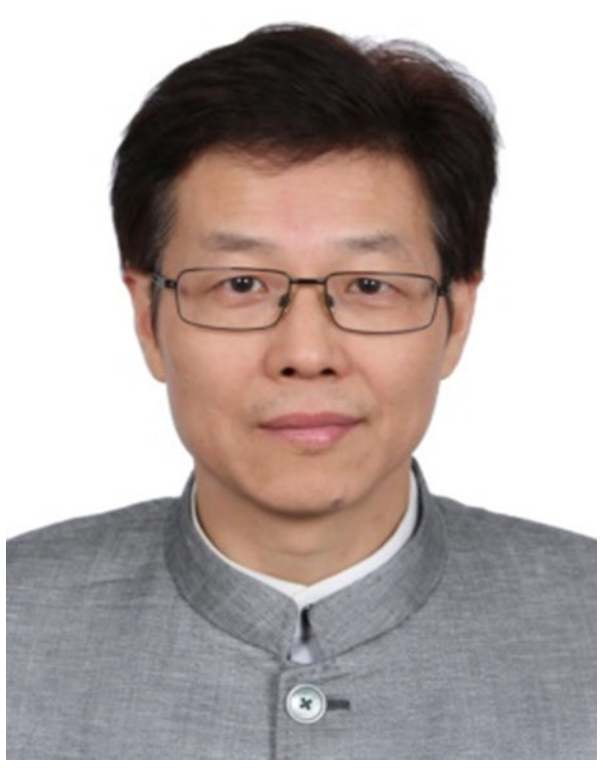

**Mingquan Lu** is a professor of the Department of Electronic Engineering, Tsinghua University, Beijing, China. He is the director of Tsinghua GNSS Research Laboratory, and a member of the Expert Group of China BeiDou Navigation Satellite System. His current research interests include GNSS signal design and analysis, GNSS signal processing and receiver development and GNSS system modeling and simulation. He received his M.E. and Ph.D. degrees in Electronic Engineering from University of Electronic Science and Technology, Chengdu, China.